\documentstyle[12pt,epsf]{article}
\def\baselinestretch{1.05}
\def\beq{\begin{equation}}
\def\eeq{\end{equation}}
\def\lleq{\label}
\def\rg{$R_{s}(p)$}
\def\fg{f_{s}(p)}
\def\za{z_{a}(s)}
\def\zh{z_{h}(s)}
\def\ns{n_{s}(p)}
\def\zd{z_{d}}
\def\dem{\delta M_s}
\def\es{E_s}

\begin{document}
\titlepage

\begin{flushright}
hep-ph/9508350 \\
HD-THEP-95-36 \\
\end{flushright}

\vspace{1 cm}

\begin{center} {\bf {SMALL SCALE STRUCTURE PREDICTIONS
FROM DISCRETE SYMMETRY BREAKING --
EARLY QUASAR FORMATION }}

\vspace{1 cm}

\rm
\vspace{2ex}
{\bf Smaragda Lola} \\
 \end{center}
 \begin{center}
{\small Institut f\"{u}r
 Theoretische Physik, Univerisit\"at Heidelberg,}\\
 {\small Philosophenweg 16, 69120 Heidelberg, Germany }\\
 \end{center}
\vspace{3ex}
\vspace{-0.5 cm}
\begin{abstract}
\noindent
We discuss the local density fluctuations
which arise due to the
topological defects that appear after the
phase transition of light pseudo-Goldstone
bosons. It has been
found that in a post-inflationary
universe the fluctuations of
these defects at large scales
may have led to galaxy formation,
while being consistent with the measurements
of the cosmic microwave background radiation.
Here we show that, at the local level, the
fluctuations may be
sufficiently large to lead to the production of smaller
structures (ie quasars)
with the observed distribution, which peaks at $z=2$
and drops rapidly for higher redshifts.
Moreover it may be possible that a limited number
of quasars are produced at redshifts of
order 10, much earlier than
what hot and cold dark matter scenarios predict.
Although in this letter we work in the parameter
space which is optimal for the generation of large scale structure as
well, these features are generic for a wide class of domain wall models.

\end{abstract}

 \noindent

 \noindent
 \rule[.1in]{14.5cm}{.002in}

\thispagestyle{empty}

\setcounter{page}{0}
\vfill\eject

\section{Introduction}

Over the recent years, there has been a
growing interest in the origin of
the observed large scale structure of the universe,
due to the data coming from COBE \cite{COBE}
and from the extensive IRAS survey
\cite{largescale}.
One of the most important
conclusions of both measurements
is that the standard cold or hot
dark matter (CDM or HDM) scenarios
with a
Harrison-Zeldovich spectrum of
primeval fluctuations,
fail to account on their own
for the complete spectrum of
energy density fluctuations.
In the case of cold dark matter,
which has been considered as the
most promising solution,
the scale invariance of the spectrum
results in a discrepancy:
either the data fits well with the theory
at large scales and then the
predicted structure at smaller scales is
unacceptably large, or
the data is normalised to agree with the
observations at small scales and
then there is not enough power at
the long wavelength components of
the theoretical
spectrum.

In several particle physics models,
there has been a lot of effort
to provide additional sources
of fluctuations at large scales
\cite{hil}
In one of these attempts,
we have
applied percolation theory in order to
perform a detailed analysis of the
density perturbations that are to be expected
from the domain walls forming
during the phase transition of very light fields
 \cite{LLOR}.
This statistical method, which has first been
introduced in the study of domain wall distributions
in the universe by Lalak, Ovrut and Thomas \cite{p2},
allowed us to formulate
a picture of the spatial
distribution of the
overdensities
in a post-inflationary
universe.
We found that
the domain walls
may act as seeds of structure
formation and enhance the
standard cold dark matter spectrum, in
such a way as to account for
the whole range of observations of IRAS and
COBE and still be consistent with the
measurements
of the cosmic microwave background radiation.
This occurs, provided that one of the minima
of the potential of the scalar field is favoured,
and in \cite{LLOR,CLB} it has been demonstrated
why this is true after inflation has taken
place\footnote{
In the case that both minima appear with the same probability,
horizon size domain walls which would result to
unacceptably large
fluctuations in the cosmic microwave background
radiation arise \cite{wal}.}.

Here,
we will extend the analysis
in the small scale distribution of matter.
In these lines, it is of particular interest
to find at what redshift the
first structures are expected to form
and how large is the
amount of mass
that has become non-linear
at that time.
The objects with the higher observed
redshift, $z_{q}$, are quasars.
In the recent years the limits
on $z_{q}$ have increased
and for the more distant quasar that has been
seen up to now,
$z_{q} \sim 5 $ \cite{quas}.
The standard
cold dark matter picture can
account for
early quasar formation
with difficulty and the
situation will become worse if
new, more distant objects are seen at even higher
redshifts. For this reason,
one would like to see whether
domain walls may trigger
sufficiently large density
fluctuations which
lead to quasar
production at early times.
This was found to be the case in \cite{p1},
where models with unstable
domain walls have been considered.
We showed that, due to a small
degeneracy in the minima of the potential
of a pseudo-Goldstone boson, which may arise
in string models, there exists a critical horizon
scale at which the true vacuum
dominates and all walls disappear.
However, when the wall bubbles that
surround a region of true vacuum
expand, rapid collisions
and domain wall annihilation occurs,
resulting in large local overdensities,
which can host quasars. Redshifts as large
as $10$ were predicted, which are much larger
than those expected in CDM models.

In schemes with stable walls, where
the field may roll to the minima of its potential
with different probability,
overdensities at the local level are
also to be expected.
Using percolation theory, it is possible
to examine not only at what
red-shifts the mass in the non-linear
regime becomes sufficient for the
formation of stellar objects, but
also what is their spatial distribution.
The data indicates
that at high and low redshifts
there exists
a decrease in the distribution of
quasars, the peak being at
$z \approx 2$.
In this work, we will attempt to
gain some insight, as to why this
occurs.

\section{Domain walls in the percolation theory picture}

Domain walls are associated
with discrete symmetries,
which arise commonly in many particle
physics models, after the
explicit breaking of a continuous
symmetry \cite{hilross}.
The resulting potential
of the pseudo-Goldstone bosons is of the form
\beq
V = V_{0} \left[ \cos \left(\frac{\phi}{\upsilon}\right) +1
\right]
\label{e1}
\eeq
and obeys the discrete symmetry
$\phi \rightarrow
\phi + 2 \pi n \upsilon$, $n = 1,2 ...$
The equation of motion corresponding to
the above potential, admits domain wall
solutions that interpolate between
two adjacent vacua \cite{walls}.
The width of the walls,
$\Delta$, is given by
\begin{equation}
\Delta = \frac{\upsilon}{\sqrt{V_{0}}} =
m^{-1}
\label{eq:e16}
\end{equation}
$m$ being the mass of
$\phi$ evaluated at any minimum.
The surface energy density of
the wall is
\begin{equation}
\sigma = \int_{-\infty}^{+\infty}
2 V_{0} \left[ \cos \left(\frac{\phi}
{\upsilon}\right) +1 \right] \, dz  = 8
\, \upsilon^2 m
\label{eq:e17}
\end{equation}

The space distribution of
the domain walls is found by
partitioning the three-dimensional
space into cubic lattice sites with
lattice spacing $\Lambda$ \cite{p2}.
In this letter, as in \cite{LLOR}, we
work with a system that has two minima.
It is assumed that at each lattice site
the physical system can be in one of the two vacua,
denoted by $(+)$ and $(-)$ respectively.
The probability that
a lattice site is in the $(+)$ vacuum
is denoted by $p$, $0 \leq p \leq 1$,
while the probability that a lattice site
is in the $(-)$ vacuum is $q=1-p$.
Provided  there is  no correlation  between the
vacuum structures at  any two different lattice
sites, it is possible to calculate
the spatial distribution of the two vacua
and, hence, the spatial distribution of
domain walls, by applying three-dimensional
percolation theory \cite{3dp}.
It has been shown that for
$p < p_{c}$, where $p_{c}=0.311$ is the
critical probability for a cubic
lattice in three dimensions,
the $(-)$ vacua lie predominantly
in a large percolating cluster (since necessarily $q>p_c$),
while the $(+)$
vacua are in finite
s-clusters.
Here, $s$ denotes the
number of nearest neighbour lattice sites that
are occupied by $(+)$.
Moreover,
it was recently found that the scaling
behaviour of the percolating cluster is
not persistent until $p$ reaches the value
$p=1/2$ \cite{CLB}.
This indicates that for a large range of
probabilities only finite wall bubbles are present.
On a given lattice, the number of s-clusters
falls rather quickly with growing s.
Indeed,
the probability per lattice site that a given lattice
site is an element of an s-cluster,
$\ns$, (which is
given by the ratio of the
total number of s-clusters, $N_{s}$, over the total number
of lattice sites, $N$) is
\begin{equation}
\ns = 0.0501 s^{-\tau}\exp \left \{-0.6299
\left(\frac{p-p_{c}}{p_{c}}\right) s^{\sigma} \left[(\frac{p-p_{c}}{p})
s^{\sigma}+ 1.6679\right]\right\}
\label{ns}
\end{equation}
where $\tau = 2.17$ and $\sigma = 0.48$ \cite{p2}.
Since for a given lattice
there exists an upper
statistical cut-off on the size of observable clusters,
no unacceptable fluctuations of
the cosmic microwave background radiation
due to domain wall bubbles are generated
\cite{LLOR}.

The mean radius of a wall bubble
at a specific redshift
is well characterized by the average radius
of gyration, \rg, of an s-cluster.
This quantity
(for $p < p_{c}$ and $s > s_{\xi}$)
is found to be
\begin{equation}
R_{s}(p) = f_{s}(p) \Lambda \equiv
 0.702 (p_{c}-p)^{0.322} s^{0.55} \Lambda
\label{Rs}
\end{equation}
where
\begin{equation}
s_{\xi} = \left(\frac{0.311}{|p-0.311|}\right)^{2.08}
\label{sz}
\end{equation}
Initially, $R_{s}$ is larger than the
horizon.
However, the horizon radius grows
faster than that of the
bubble (whose radius just
grows linearly with the expansion),
and thus at some
redshift $\za$ the bubble comes
within the horizon.
At this stage the bubble
shrinks under its surface tension,
undergoes a few cycles of oscillations,
and finally
looses its energy in the form of scalar waves
\cite{WWW}.
The energy stored in the domain wall is
\beq
E_{s} = f t_{s} \sigma \Lambda^{2}, \; \; \; t_{s} =
s \left ( \frac{1-p}{p} \right )
\label{col}
\eeq
where $t_{s}$ is the average surface area of an
s-cluster.
The parameter $f$ is $1 \leq f \leq 6$ and
here we use the moderate value  $f=3$.
Due to the expansion of the
universe, the lattice spacing $\Lambda$ is a function of
the redshift $z$.
The redshift
when the wall bubbles shrink, $\za$,
is taken to be the redshift
when an s-cluster enters
the horizon. The overdensities that are
expected to arise after the bubbles shrink are calculated in a
subsequent section. However, even at this stage, it is
possible to predict the qualitative picture that arises
and to see its relevance for quasar formation.

\section{Quasar production and spectrum}

Quasars are high-redshift active galaxies,
with a very energetic central source of energy,
which may not be coming from nuclear fusion.
The most popular scheme is that quasars are powered
by the accretion of matter in a supermassive
black hole ($M_{h} \approx 10^{8} M_{\odot}$)
in the center of a host galaxy
\cite{bookqua}. Then, before any quasar activity can
begin, some galaxies must have formed and
virialised at redshifts higher than that of the
actual quasar and subsequently evolved
to the stage of developing a
massive black hole.
This indicates that the very existence of
quasars implies that non-linear structures
must have appeared at redshifts higher than
$5$.

Studies of the space distribution of quasars show
that their number density
exhibits a peak at a redshift
$z \approx 2$; for smaller redshifts the observed
abundance of
quasars decreases.
The space density of quasars with redshift $z < 2$
has been measured to be roughly
$10^{-5} \; {\rm Mpc}^{-3}$.
For redshifts $z > 2$ there is
also a decrease of the number of quasars as the redshift
goes up,
except for the very bright ones. This decrease however
is gradual, rather than a steep cut-off.
It is also possible to  estimate the mass
that ought to collapse at a redshift $z_{q}$, in order
to lead to quasar production at a later time.
For example, for a quasar with a
redshift $4$
it is found that the minimal collapsing mass should be
at least $O(10^{12} M_{\odot})$,
and this value can be even larger.

However, such a picture for the local
fluctuations is exactly
what we would expect from percolation theory.
The basic points to note are that:

$\bullet$
the larger domain wall bubbles
enter the horizon and shrink under
surface tension later than the smaller ones.
Since

$\bullet$
wall-driven fluctuations redshift
slower than radiation or matter,

$\bullet$
the local fluctuations that involve larger
bubbles may become non-linear at higher redshifts.
On the other hand,
percolation theory predicts that

$\bullet$
the number of lattice sites
$n_{s}$ decreases exponentially with $s$,
thus the number density of bubbles decreases
with their size.

The combined effect is that
{\em non-linear fluctuations
at very high redshifts
become rare}.
As for the decrease in the
number of quasars as the redshift
drops below $2$ it can also be explained.
This is because

$\bullet$
For larger wall bubbles, the amount of
mass in the non-linear regime is
larger. It is possible therefore that
at $z = 2$ we approach a critical mass scale which is
slightly higher from the minimum value that we need in order for
quasar formation to proceed.
Then, at that redshift the quasar
density will be expected to be maximal, while
at lower redshifts the number of objects
will decrease.

$\bullet$
Finally, for the brighter quasars
more mass is required
in the non-linear regime,
therefore they will appear only at high redshifts.

Note that these features are generic and lead
to the same qualitative picture for
different regions of the parameter space and in particular for
different choices of the mass of the field, $m$, which
sets the time-scale of the transition\footnote{
Only when the length scale $m^{-1}$ comes within the horizon,
can the field roll towards its minima \cite{LLOR}
.}.

\section{Local density fluctuations}

While the qualitative behaviour that has been discussed is generic,
in order to gain a better understanding, we will calculate the local
density fluctuations in a specific scheme. Here, we will chose to
work with the same parameter space which in  \cite{LLOR} was found
to lead to the best agreement with the data at large scales.

In \cite{LLOR}, in order to compare the
wall driven fluctuations with the ones observed at the
large scales, the energy density
$E_{s}$ had been averaged over the
mean distance between s-clusters
at $z_{a}$.
The Fourier analysis of a quasi-periodic matter distribution
shows that the amplitude of the
Fourier coefficients is peaked in the
momentum space around a set of discrete
points corresponding to the wavelengths
$\lambda =\infty,d,d/2,...,R$,
where $d$ is the mean distance between seeds
and $R$ is the  typical radius of the overdensity
produced by the accretion of matter onto the seed
in question. In the scenario described in \cite{LLOR}
it has been assumed that $R \approx d$
(that is a big amount of the
overdensity detached by
each cluster is dispersed over a
region not much smaller than $d$).
This is a reasonable assumption as the seeds
considered in \cite{LLOR} are produced
in the radiation dominated epoch,
and one can argue that the accretion is not very effective at that time.
However, we do not know how large $R$ really is.
One may expect that the larger clusters, which {\em enter the
horizon at a much later stage}, may
drive collapse more efficiently and leave behind more compact
overdensities whose radii are smaller than $d$  and as small as $R_s$.
We think that this is a reasonable assumption,
for the following reasons:

$\bullet$
Although the phase transition occurs deep in the radiation
dominance, the larger clusters enter
the horizon at the very end of
the radiation era. At that time matter has already started
to slow down significantly, therefore part of it may
accrete around the wall just before the latter collapses.
In this case we can define an ``efficiency parameter'',
$\gamma_{\ell}$,
which scales with the redshift. This parameter should become
unity in matter dominance (where subsequently
the fluctuations grow proportionally to the redshift),
while the earlier the wall bubble enters the horizon,
as compared to the beginning of the matter dominance
era, the smaller the efficiency parameter becomes.
We come back to this point in the quantitative examples
that will be presented.

$\bullet$
The larger walls store more energy, therefore they
drive more efficient collapse.

$\bullet$
In addition, although in \cite{LLOR} we have
prefered to work with non-interacting
light scalar fields, as they are out of equilibrium and
their mass is naturally in the correct range for
structure formation, it is also
possible that  some {\em interaction} between the
light fields and ordinary matter is present.
In \cite{mh} it has been shown that very light pseudo-Goldstone
bosons may give rise to long range forces.
In the case that an interaction exists, localized density
fluctuations due to domain walls may appear
\cite{masarotti}. If this interaction is very weak,
it still will not be sufficient
to change the out-of-equilibrium property of the system,
which has the general behaviour that we have described
in \cite{LLOR}.
Nevertheless, even a very soft interaction can result in
amplifying the local energy dissipation
mechanism in comparison to the situation where only gravity is present
in the theory. As we are going to see below, an
efficiency for matter accretion as low as $ \approx
10\%$, around collapsing
domain walls which enter the horizon at the end of the radiation era,
is sufficient to support the picture we propose.

$\bullet$
Finally, while here we chose to work with the parameter space that
was found in \cite{LLOR} to be optimal for the creation of
large scale structure due to a single phase transition, it
would be possible to relax this condition. In such a case, we
can assume that the mass of the pseudo-Goldstone boson, $m$,
can be smaller, such that the larger walls which
give rise to the local density fluctuations that subsequently
will host quasars, enter the horizon and dissipate their energy
in the matter dominance era.
In \cite{LLOR} $m$ was fixed
by demanding that the peak of the density fluctuations as
a function of the scale occurs at $\approx 30$ Mpc, as
observations indicate. What happens when $m$ is smaller?
The first point to make is
that the density fluctuations at large scales will
decrease, since the fluctuations now have less time to grow
(here we should note that for large scale structure, the relevant
fluctuations are of super-horizon size and grow
as the square of the red-shift during the radiation era
\cite{KT}, while the local fluctuations practically grow only
in matter dominance).
The larger domain walls may now enter
the horizon deep inside
matter dominance and the resulting local
density fluctuations
are amplified. As we have pointed out,
the basic features which determine the quasar
distribution {\em are generic} and give rise to a similar
qualitative behaviour if a smaller $m$ is chosen.

The picture we have therefore
is that overdensities
at scales larger than the respective $R_s$,
as well as the local overdensities which do not become nonlinear
early enough, form a kind of a diffusive
background on top of which some
overdensities at the local level
may form gravitationally
bound structures, for instance galaxies
hosting quasars.
In what follows we will try
to estimate the expected spectrum of such structures.
For this, we need to calculate the overdensity at a local
level and the scale over which the averaging is done is
the one that may provide mass greater than
this of a quasar,
in the non-linear region $\delta \rho/\rho \geq 1$.
The smallest possible distance
over which we may average
(leading to the larger local
fluctuations) is the horizon
at a redshift $z_{a}$,
$R_{H_{a}} \equiv
R_{H}(z_{a})$.

Let us now pass to specific formulas:
The redshift $z_{a}$ is obtained
by equating the mean radius of gyration
for an s-cluster,
to the horizon at that redshift.
We find that
\beq
1+z_{a} = \frac{1+z_t}{\alpha  \fg}
\eeq
with
\beq
(1+z_t)^2 = \frac{R_{{H}_0}} {R_{H}(z_t)}
(1+\zd)^{1/2}
\eeq
where $z_d$ is the redshift when
matter domination begins
and $R_{H_{0}}= 6000$ {\rm Mpc}
is the horizon size today\footnote
{Throughout the calculation we are going to
assume that the reduced
Hubble constant $h$ is unity (that is the
Hubble constant today is
$H_{0} = 100 \,{\rm km} \,
{\rm s}^{-1} \,
{\rm Mpc}^{-1}$), for
simplicity of presentation. A different value of
$h$ does not alter the picture we have. In this case,
the input model parameters that are needed to
fit the large scale data, which are
the same that we use here for small scale
predictions, are shifted to
$\sigma h^2$ and $\upsilon h$,
as explained in \cite{LLOR}
.}. The factor $\alpha \equiv
H(z_t)/H(z_f)$, where
$z_t$ is the redshift where the field
starts rolling down the potential towards
one of its minima, and $z_f < z_t$ denotes the time
 at which the system actually settles in one of the vacua,
after a period of oscillations \cite{LLOR}.
The above formulas hold for
$z_{a} \geq z_{d}$.

Assuming
that the mean total energy density
is equal to the critical
density, we may express the
critical energy density at $z_{a}$ in
terms of the present day critical density $\rho_0$
as
\beq
\rho_c (z_{a}) = \rho_0 \, \frac{(1+z_{a})^4}{1+\zd}
\eeq

The local energy density perturbation due to an
s-cluster with diameter $R_{H_a}$ at $\za$ is
\beq
\left. \frac{\delta \rho}{\rho}
\right|_{a} \equiv
\left. \frac{\delta \rho}{\rho}
\right|_{\ell oca\ell} (z_{a})
= \frac{ 6 \, f \, \sigma \,
(1-p) \, s \, \Lambda_{a}^{2}}
{p \, \rho_{c}(z_{a}) \,
\pi \, R_{H_{a}}^{3}}
\lleq{loc}
\end{equation}
where
\beq
\Lambda_{a} \equiv
\Lambda(z_{a}) = \frac{\alpha}{m} \,
\frac{1+z_{t}}{1+z_{a}}
\lleq{loc1}
\eeq
is the lattice spacing at $z_{a}$\footnote{ $\Lambda(z_t)
\equiv a/m$ \cite{p2,LLOR}.}
and
\beq
R_{H_{a}} = \frac{1}{m} \, \left
(\frac{1+z_{t}}{1+z_{a}} \right)^{2}
\lleq{loc2}
\eeq

The local
fluctuations are
(in contrast to the fluctuations that give rise
to the large scale structure)
always sub-horizon and therefore grow
logarithmically with the redshift during radiation
dominance and linearly during matter dominance.
Then the redshift $z_{q}$ at which the
fluctuations become
non-linear is given by
\beq
1+z_{q} = \gamma_{\ell} \;
\left.
\frac{\delta \rho}{\rho}
\right|_{a}
(1+z_{d}) \left(
1 + 2 \log \frac{1+z_{a}}{1+z_{d}}
\right)
\lleq{loc3}
\eeq
The factor $\gamma_{\ell} < 1$ which
appears in (\ref{loc}), has been added
in order to take into account that for the parameter
space where we work, at
the end of the radiation dominance
only a part of the overdensities
that are produced by the wall bags
will remain localized.

The total amount of mass
in the non-linear region at $z_{q}$,
$M_{q}$, in terms of solar masses $M_{\odot}$,
is given by
\beq
M_{q} = \frac{ \pi L_{q}^{3}}{6 M_{\odot}}
\rho_{c}(1+z_q)^{3}
\lleq{loc4}
\eeq
where the scale of the perturbation
at $z_{q}$ is
\beq
L_{q} = R_{H_{a}} \, \frac{1+z_{a}}{1+z_{q}}
\lleq{loc5}
\eeq
and $M_{\odot}$
is the solar mass.
Finally, we can identify the space distribution of
the local fluctuations.
The average distance between s-clusters at a
redshift $z$ is
\beq
d(z) = \left(\frac{V(z)}{V(z)n_{s}}\right)^{1/3}
\Lambda(z)
\lleq{loc6}
\eeq
thus today
\beq
d(z=0) = \frac{1}{n_{s}^{1/3}} \frac{a}{m}
(1+z_{t})
\lleq{loc7}
\eeq

\section{Numerical analysis}

In \cite{LLOR} we have found that
large scale structure may form
as a result of the global density fluctuations
(fluctuations averaged on scales $d$),
and some indicative
combinations for $\alpha =10$
appear on Table 1.
Here we want to use the same set of parameters
to examine the local overdensities of the model.
However, in \cite{LLOR}
we had also introduced a parameter
$\gamma_{s}$  (to account for the
fact that before a wall bag disappears,
it may stay around sufficiently long
to cause the collapse of amounts of matter).
This coefficient may not be determined
precisely without a more detailed analysis
in the framework of the spherical collapse model.
In \cite{LLOR} we took the value $\gamma_{s} =10$, however
if this parameter is of order unity,
the only modification in our
results would be that we need a higher value of
$\upsilon$ to fit the IRAS and COBE data. The
model parameters that lead to solutions,
for $\gamma_{s} = 1$,
are given in Table 2.

In the present work
we take $\gamma_{s}$ at its minimum value and
on top of that we have introduced
$\gamma_{\ell}$, to account for the fact that
the wall bubbles that we consider appear
in the end of the radiation dominance (if $\gamma_{s} > 1$, even larger
efficiency in
the accretion of the local overdensities would
be expected).
To see whether it is possible to
get the correct qualitative behaviour
for the distribution of local overdensities,
we consider the following possibilities:

(i) $\gamma_{\ell}$  $O(0.2)$.
Such a constant factor (especially
shifted towards larger values)
may be expected if some
soft interactions are present.

(ii) $\gamma_{\ell}$  $O(0.1)$.

(iii) In the absence of interactions, the
most realistic approach is
to take into account
that larger bubbles
lead to a more efficient energy dissipation,
since they enter the horizon nearer the matter dominance era,
where the overdensities grow linearly with the
redshift.
For this reason we set
$\gamma_{\ell} = (1+z_d)/(1+z_a)$. For
$z_{a} \approx z_{d}$ the efficiency parameter
is unity, since we are in matter dominance,
while the higher $z_{a}$ is, the smaller the
parameter becomes.

Using the model parameters of Table 2, we
have calculated the local density fluctuations
as well as their space distribution,
the redshift $z_{q}$ where the
fluctuations become non-linear and
the amount of mass
that is involved in the non-linear regime.
The results appear on
Tables 3,4 and 5
for the three choices of
input parameters respectively.
These tables indicate that
for all three choices,
the amount of mass in the non-linear region
can be sufficiently large to allow for
early quasar formation.
Moreover, we reproduce
qualitatively the observed
space distribution
of quasars, at redshifts $z \geq 2$,
that is quasars at
larger redshifts appear with
larger space separation.
We also observe that the mass in the non-linear
regime reduces with the red-shift, indicating that
after a certain redshift the total available amount of
mass will be near the lower limit that we
need for quasar production. We find that for this
to occur at $z =2$, the mass should be
$O(10^{13} \; M_{\odot})$.
Concerning the scale of the perturbation,
in all cases is found to be $O({\rm Mpc})$.
We also see that the number of quasars at a specific redshift
is sensitive to the parameter $p$.
Indeed, for p = 0.11 we find that $s = 50$
leads to one quasar every 1614 Mpc, while
for p =0.15 the distance is 461 Mpc.

The number of quasars and the total mass
in the non-linear regime, as functions of the red-shift,
for the three cases of Table 2,
are given in figures 1-6.
We see that the qualitative behaviour is in agreement
with observations and that even with the
rough approximations that we have made,
the quantitative agreement is also good.

\section{Conclusions}

To summarise,
we have looked at the local density fluctuations
generated by domain walls
after the phase transition of light pseudo-Goldstone
bosons. In particular, we have analyzed the
expected density perturbations and
their spatial distribution, as well as
the redshifts at which they become non-linear.
We have found that, complementary to the generation of the
observed large scale structure,
the same overdensities
may lead at the local level to an early
appearance of non-linear
fluctuations which may result to
early quasar production.
The scale of the overdensities
is naturally of the correct order of magnitude.
Concerning the spectrum of these objects, we
show that
quasars are expected to
appear with larger space separation
as the redshift increases, in consistency with
observations. A decrease to the number of objects as
the redshift falls below a critical value is also
predicted.
The total amount of mass that is involved in
this non-linear process is
from $10^{12}-10^{14} M_{\odot}$, which is interesting, given
that $10^{12} M_{\odot}$
is the minimal possible value for
an overdensity to evolve to a galaxy that may
host a quasar.

\vspace{0.2 cm}
\noindent
{\bf Acknowledgment }

\noindent
I am grateful to Z. Lalak,
B. Ovrut and G. G. Ross for many
enlightening discussions
on the subject.

\pagebreak

\noindent{\bf Table Captions}

\vspace{0.4 cm}

{\bf Table 1.}
 Model parameters generating
the observed large scale structure that were presented
in \cite{LLOR} (where amplification of the fluctuations
due to matter accretion in the wall before the later
collapses were considered).

\vspace{0.3 cm}

{\bf Table 2.}
 Model parameters
generating the observed large scale structure
for minimal accretion of matter in the
wall before the later collapses.
The only change from table 1 is in the parameter $v$.

\vspace{0.3 cm}

{\bf Table 3.}
 Astrophysical parameters of the quasars
for case 1 of table 2. The redshifts
$z_{q_{1}}$, $z_{q_{2}}$, $z_{q_{3}}$
stand for the three possibilities for
the efficiency parameter,
$\gamma_{\ell} = 0.2$,
$0.1$ and $(1+z_d)/(1+z_a)$
respectively. In this table
as well as in the following ones
we stop the calculation as
soon as $d$ grows beyond the horizon today.

\vspace{0.3 cm}

{\bf Table 4.}
 Astrophysical parameters of the quasars
for case 2 of table 2.

\vspace{0.3 cm}

{\bf Table 5. }
Astrophysical parameters of the quasars
for case 3 of table 2.

\vspace{0.8 cm}

\noindent{\bf Figure Captions}

\vspace{0.4 cm}

{\bf Figure 1.}
The number of quasars as a function of the
redshift, for $\gamma_{\ell} = 0.2$.
The symbols $+$, $*$ and $\times$ correspond to the three cases
of Table 2 respectively. The notation is the same
in the rest of the figures as well.

\vspace{0.3 cm}

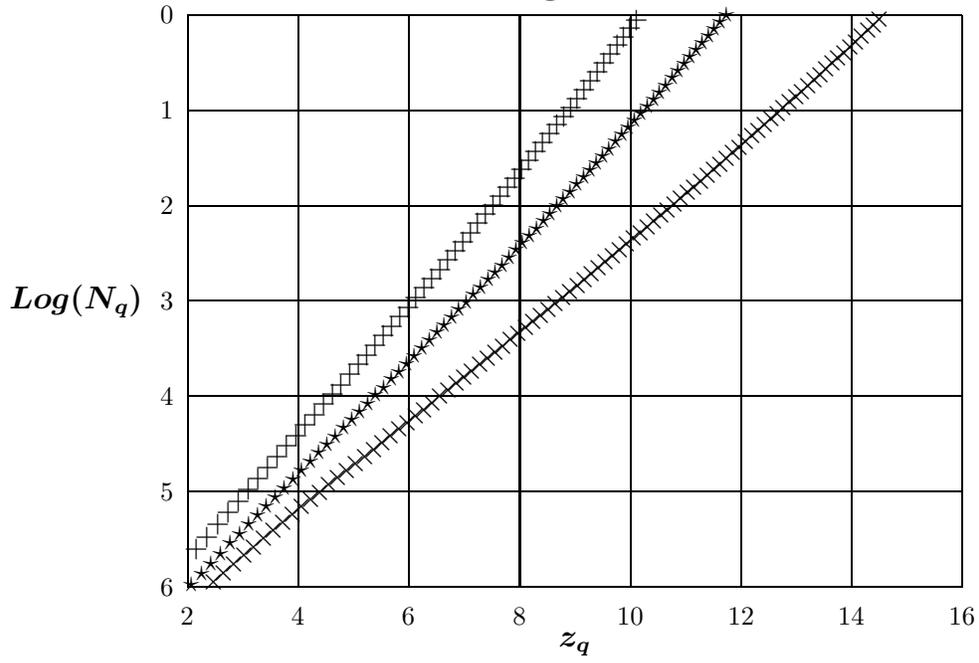
{\bf Figure 2.}
The mass in the non-linear regime as a function
of the
redshift, for $\gamma_{\ell} = 0.2$.

\vspace{0.3 cm}

{\bf Figure 3. }
The number of quasars as a function of the
redshift, for $\gamma_{\ell} = 0.1$.

\vspace{0.3 cm}

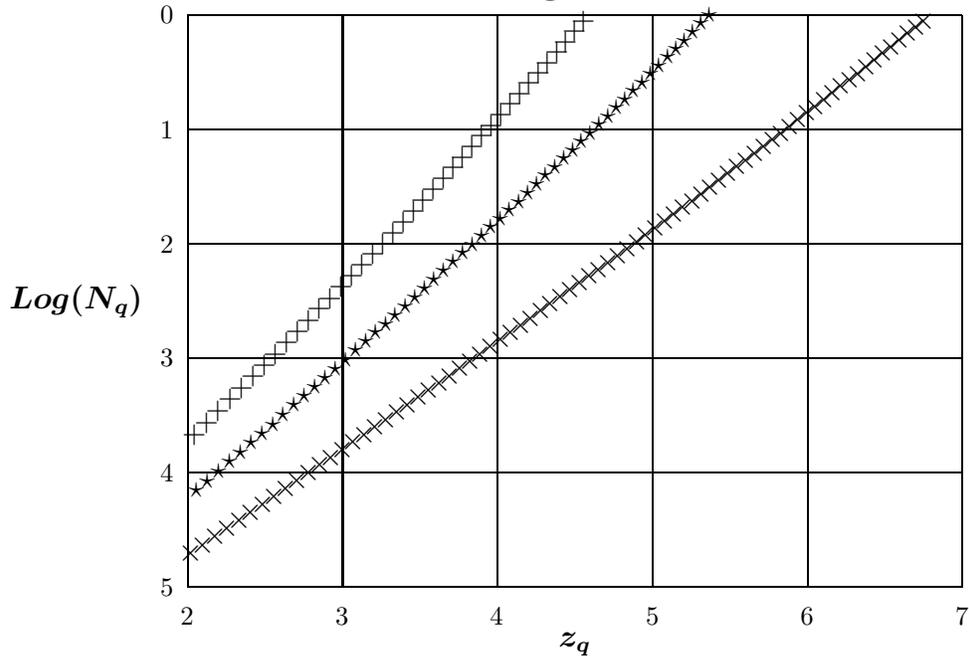
{\bf Figure 4. }
The mass in the non-linear regime as a function
of the
redshift, for $\gamma_{\ell} = 0.1$.

\vspace{0.3 cm}

{\bf Figure 5. }
The number of quasars as a function of the
redshift, for $\gamma_{\ell} = (1+z_d)/(1+z_a)$.

\vspace{0.3 cm}

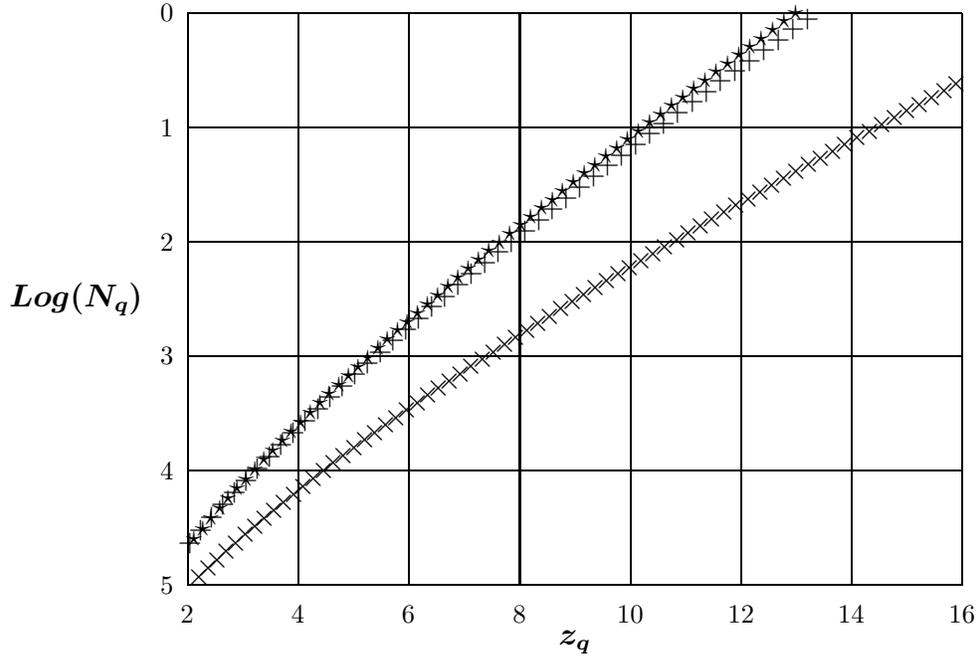
{\bf Figure 6. }
The mass in the non-linear regime as a function
of the
redshift, for $\gamma_{\ell} = (1+z_d)/(1+z_a)$.

\pagebreak

\begin{table}[t]
\centering


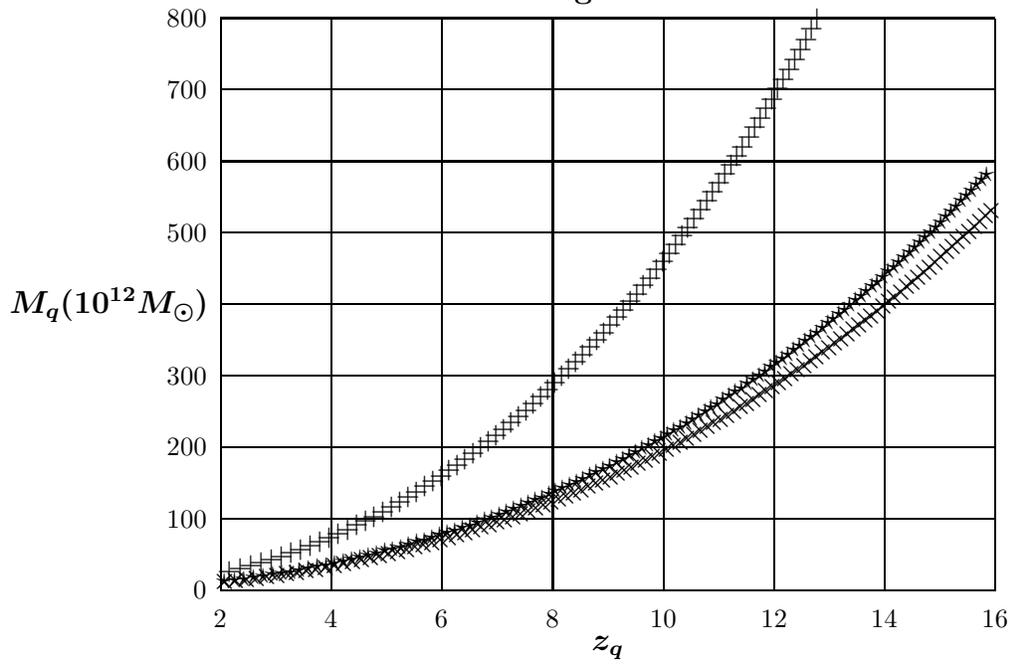
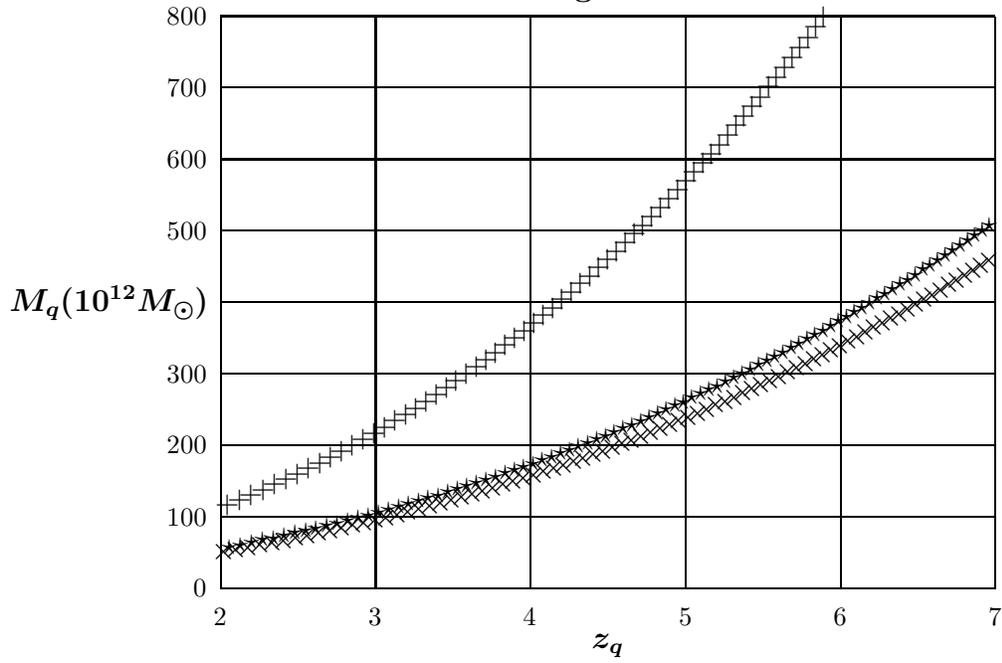
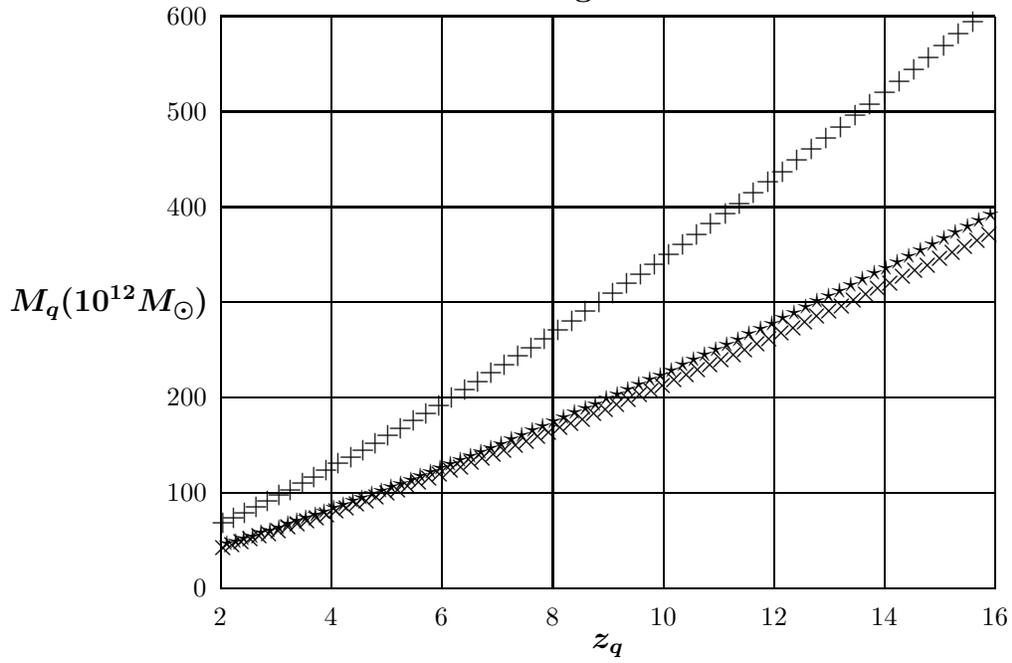
\end{document}